\documentclass[aps,twocolumn,floats,showpacs,superscriptaddress]{revtex4}
\pdfoutput=1
\usepackage{graphics,dcolumn,float}
\usepackage{graphicx}
\usepackage{dcolumn}
\usepackage{bm}
\usepackage{amsmath}
\usepackage{amssymb}
\usepackage{color}
\usepackage{soul}
\begin{document}
\preprint{APS/123-QED}
\title{Normal electric field enhanced light-induced polarizations in silicene}

\author{N. Shahabi}
\email[Corresponding author's Email: ]{shahabi@azaruniv.ac.ir}
\affiliation{Department of Physics, Azarbaijan Shahid Madani
University, 53714-161, Tabriz, Iran}
\affiliation{Condensed Matter
Computational Research Lab. Azarbaijan Shahid Madani University,
53714-161, Tabriz, Iran}
\author{A. Phirouznia}%
\affiliation{Department of Physics, Azarbaijan Shahid Madani
University, 53714-161, Tabriz, Iran}
\affiliation{Condensed Matter
Computational Research Lab. Azarbaijan Shahid Madani University,
53714-161, Tabriz, Iran}

\date{\today}
\begin{abstract}
The role of staggered potential on light-induced spin and pseudo-spin polarization has been investigated in silicene. It has been shown that non-equilibrium spin and pseudo-spin polarizations are emerged in silicene sheet by applying an external perpendicular electric field in the presence of circularly polarized light emission. This electric field results in pseudo-spin resolved states very close to the Dirac points therefore could be considered as a pseudomagnetic field. It has been shown that staggered potential induced spin-valley locking and pseudo-spin resolved bands are responsible for enhancement of the spin and pseudo-spin polarizations. Meanwhile, spin-valley locking suggests a coexistence of both spin and valley polarizations with nearly identical population imbalance at low Fermi energies which could be employed for magnetic detection of the valley polarization. It has been shown that spin-valley locking results in protection of the spin polarizations against the relaxations in elastic scattering regime. In addition results indicate that the pseudo-spin current can be generated by the circularly polarized light which could be explained by asymmetric light absorption of the states in $k$-space.
\end{abstract}


\maketitle

\section{INTRODUCTION}
Linear dispersion relation in the vicinity Fermi energy in graphene leads to semi-metallic behavior which is described by the massless Dirac theory \cite{geim2010rise}. Graphene shows a variety of outstanding electronic and optical properties which make it one of the best candidates of optoelectronic applications \cite{bonaccorso2010f}.

Monolayer graphene-like materials have attracted considerable attention during last decade. They inherit their rich physics from their famous counterpart i.e. graphene. Analogous to graphene, the basic structure of graphene-like materials is honeycomb lattice. The physics of this group of two dimensional (2D) materials is described by the massive Dirac fermions theory which stems from rather large intrinsic spin-orbit coupling that the latter originates from structural buckling  \cite{ezawa2012valley}. The large ionic radius of silicon and group IV elements of the periodic table, leads to this structural buckling \cite{ezawa2012valley}.

Another prominent feature of buckled structures is the tunable band gap that can be controlled by an external electric field  \cite{ezawa2015monolayer}. Silicene, germanene and stanene illustrate many attractive electronic and spintronic properties. Biased single layer silicene and germanene were reported to work effectively as field effect transistors while a vertical electric field can open a band gap in their semi-metallic band structure \cite{ni2011tunable}.

Edge manipulation of the mentioned 2D materials were studied which can suggest very promising applications: a giant magnetoresistance which can lead to a “topological quantum transistor” and a perfect spin filter  \cite{rachel2014giant}.
Potential application of silicene, germanene and stanene for Na or Li ion storage in Na or Li batteries has also been investigated \cite{mortazavi2016application}.

Any two-component quantum degree of freedom which is mathematically equivalent to spin, can be considered as ‘pseudo-spin’. In 2D systems with two sublattices, pseudo-spin portraits the sublattice degree of freedom with the eigenstates localized on $A$ or $B$ sublattices \cite{pesin2012spintronics}. Pseudo-spin is analogous to the true electron spin but behaves totally different under time-reversal and parity inversion \cite{trushin2011pseudospin}. Pseudo-spin stems from the degeneracy between two inequivalent atomic sites per unitcell \cite{mecklenburg2011spin}. Meanwhile, \cite{mecklenburg2011spin}, Mecklenburg et al have demonstrated that the sublattice state vector represents indeed a real angular momentum in $3+1$ dimensions.

Pseudo-spin is not associated with the internal magnetic moment of electron and does not interact with an external magnetic field however, can manifest itself in observable quantities and can be detected in transport phenomena and inter-band optical absorption \cite{trushin2011pseudospin}. Trushin et al have shown that due to the out-of-plane orientation of pseudo-spin, switching the helicity of circularly polarized light can cause a reduction or enhancement in inter-band absorption because the elements of inter-band transition matrix are sensitive to the light polarization and also pseudo-spin orientation in the initial and final states \cite{trushin2011pseudospin}. Although the pseudo-spin eigenstates are really robust, the exchange electron-electron interaction can alter the pseudo-spin orientation \cite{trushin2011pseudospin}. 

Interesting effects arise as a result of the buckling in silicene therefore particular interest of the present research is focused on this material. Among different proposed hybridizations, buckled silicene, which would be named as Si(111), is described by the hybridization in the second-nearest neighbor (2NN) tight-binding model. This hybridization is the most desired one due to stability reasons \cite{matusalem2015stability,guzman2007electronic,PhysRevB.88.035432,PhysRevB.72.075420}.  Orbitals form  bonds with their second nearest neighbors (2NNs). These bonds are responsible for conduction of the buckled silicene (Si (111)).

In this work, light-induced non-equilibrium pseudo-spin polarization has been studied. Results show that pseudo-spin polarization and also pseudo-spin-polarized current are injected in silicene due to the external perpendicular electric field which determines the silicene phase. Applying this field leads to inversion symmetry breaking. 

Another intriguing feature in 2D systems is spin-valley locking. Two inequivalent energy extrema in energy dispersion of these materials introduce ‘valley’ degree of freedom which is very promising in the new research field; ‘vallytronics’ \cite{tao2019two}. The valley polarization is an imbalance in the electron population in two valleys \cite{tao2019two}. The systems which exhibit three following features: Preserving P and T symmetries; owning two well-distinguished energy extrema called valleys which are transformed to each other by T symmetry; and having large spin-orbit coupling; are good candidates for injecting valley-dependent spin polarization by an external electric field. Before applying this electric field, system is spin degenerate and also valley degenerate. By applying electric field, which is responsible for P symmetry breaking, large spin-orbit coupling in silicene leads to spin sub-bands splitting as if they were exposed to a magnetic field. Therefore this effect is called Zeeman-like splitting \cite{tao2019two}. As no external magnetic field is applied, the system is T-symmetric. Therefore the opposite spin polarizations are formed in two valleys. This spin polarization emerged in valleys is electric-field reversible \cite{tao2019two} i.e. changing the sign of electric field, which transforms two valleys to each other, will lead to flip of spin polarization. In spite of lack of knowledge about spin and valley relaxation timescales, it has been reported that spin-valley locking effect enhances spin and valley relaxation times \cite{xu2014spin,tao2019two}. Another interesting feature of this effect is the ability of controlling the valley polarization via lifting the valley degeneracy \cite{tao2019two}. In the present study it has been shown that spin-valley locking provides a powerful framework for magnetic-based valley polarization measurement. In addition it has been demonstrated that the inversion symmetry breaking normal electric field enhances the light induced pseudo-spin polarization and pseudo-spin current significantly.   
  
The rest of present paper is organized as follows: Sec. II represents the system characteristics briefly. Sec. III explains the considered model to describe silicene. Sec. IV is devoted to the significance and role of staggered potential in silicene physics. In Sec. V the previous works in the field generation and detection of valley polarization has been summarized and Sec. VI provides discussion and results. Finally, present work is summarized and a conclusion of main results are given in  Sec. VII.
\section{SYSTEM CHARACTERISTICS}\label{system}
\begin{figure}[h]
	\includegraphics[width=1.0\linewidth]{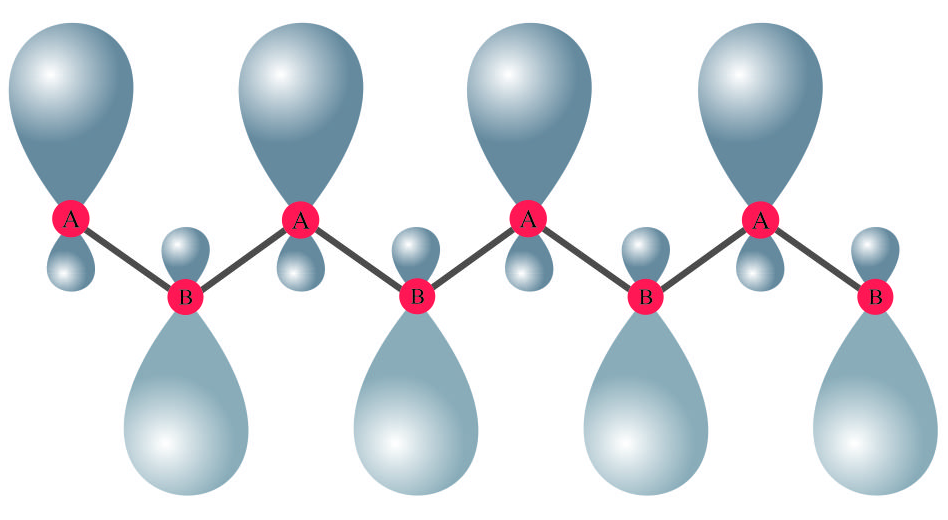}
	\centering
	\caption{(Color online) Schematic side view of the silicene $\pi $ orbitals in buckled configuration. }
\end{figure}
In order to get $\pi$-bonds of $sp^3$ orbitals along the $z$-axis we choose a linear combination of $s$, $p_x$ , $p_y$ and $p_z$ in which $sp^3$ orbitals take the forms below:

\begin{eqnarray}
\phi_{sp^3}^{(1)}(\theta ,\varphi )&=&\frac{1}{2}s(\theta ,\varphi )+\frac{\sqrt{3}}{2}p_z(\theta ,\varphi )\\
\phi_{sp^3}^{(2)}(\theta ,\varphi )&=&\frac{1}{2}s(\theta ,\varphi )-\frac{1}{\sqrt{12}}p_z(\theta ,\varphi )+\sqrt{\frac{2}{3}}p_x(\theta
	,\varphi )\nonumber\\
\phi_{sp^3}^{(3)}(\theta ,\varphi )&=&\frac{1}{2}s(\theta ,\varphi )-\frac{1}{\sqrt{12}}p_z(\theta ,\varphi )-\frac{1}{\sqrt{6}}p_x(\theta
,\varphi )\nonumber\\&&
+\frac{1}{\sqrt{2}}p_y(\theta ,\varphi )\nonumber\\
\phi_{sp^3}^{(4)}(\theta ,\varphi )&=&\frac{1}{2}s(\theta ,\varphi )-\frac{1}{\sqrt{12}}p_z(\theta ,\varphi )-\frac{1}{\sqrt{6}}p_x(\theta
	,\varphi )\nonumber\\&&-\frac{1}{\sqrt{2}}p_y(\theta ,\varphi )\nonumber	
\end{eqnarray}
$\phi_{{sp}^3}^{(1)}$ contributes in $\pi$-bonds while $\phi_{{sp}^3}^{(2)}$, $\phi_{{sp}^3}^{(2)}$ and $\phi_{{sp}^3}^{(4)}$ contribute in the $\sigma$-bonds.	
Considering only the orbital which form $\pi$-bonds, one can write
\begin{eqnarray}
\phi_{{sp}^3}^{(1)}(r,\theta ,\varphi )
&=&\frac{1}{2}R_{30}(r)Y_{00}(\theta,\varphi)+\frac{\sqrt{3}}{2}R_{31}(r)Y_{10}(\theta ,\varphi),\nonumber\\
\end{eqnarray}
in which $Y_{lm}(\theta,\phi)$ and $R_{nl}(r)$ are the spherical harmonics and atomic radial wave functions respectively (see appendix \ref{Ylm}-\ref{Rnl}). 
\\

Overlap of parallel  orbitals on adjacent Si atoms which are normal to the structure plane form $\pi$ bonds in silicene. $\phi_{{sp}^3}^{(1)}$ orbitals of $\pi$-bonds are z-oriented and have the key role in electron and spin transport. The other $sp^3$ orbitals take part in $\sigma$-bonds which are covalent-types bonds and have no direct influence on charge transport.

Using these $\pi$-bond orbitals one can develop the tight-binding approach as described in the next section.  
\section{Model}
More or less analogous to graphene, monolayer graphene-like materials are also honeycomb lattice structures and can be described by the tight-binding model \cite{ezawa2013m,farajollahpour2018anisotropic}.
The rather strong intrinsic spin-orbit coupling is responsible for spin dynamics \cite{inglot2015m} and makes graphene-like structures topological insulators \cite{ezawa2015monolayer}.
In addition, there is a layer separation between the two sublattices in their buckled structure which makes the gap tunable by applying a  perpendicular electric field, $E_z$ \cite{ezawa2015monolayer}.
Therefore the Hamiltonian describing graphene-like materials takes the following form \cite{ezawa2013m,farajollahpour2018anisotropic}. The values of parameters vary for different graphene-like structures. Hamiltonian of the silicene is given by the following expression:
\begin{eqnarray}
H=H_0+H_{SO}+H_{intR}+H_{extR}+H_b
\end{eqnarray}
\begin{eqnarray} 
H&=&-t\sum_{\langle {ij}\rangle \alpha } c^{\dagger }_{{i\alpha}}c_{{j\alpha }}+it_{{SO}}\underset{\langle \langle {ij}\rangle \rangle \alpha \beta }{\sum u_{ij}}c^{\dagger}_{{i\alpha}}\left(\sigma ^z\right)_{\alpha \beta }c_{{j\beta}}\nonumber\\&&-{it}_{{intR}}\underset{\langle \langle {ij}\rangle \rangle \alpha	\beta }{\sum u_{\text{ij}}}c^{\dagger }_{{i\alpha }}\left(\overset{\rightharpoonup }{\sigma }\times \overset{\rightharpoonup }{d}_{{ij}}\right)^z_{\alpha
		\beta }c_{{j\beta }}\\&&+
it_{{extR}}\sum_{\langle {ij}\rangle \alpha \beta } c^{\dagger }_{{i\alpha }}\left(\left({\sigma}{d}_{ij}\right)^z\right)_{\alpha \beta }c_{{j\beta }}+l\sum_{i\alpha} \zeta_iE_z^ic^{\dagger }_{{i\alpha }}c_{{j\alpha }}\nonumber
\end{eqnarray}
Where $c_{i\alpha}^\dagger$ 
creates an electron with spin polarization $\alpha$ and orbital state of $\phi^{(1)}_{sp^3}$ at site $i$ and $c_{i\alpha}$ annihilates 
an electron with spin polarization $\alpha$ and orbital state $\phi^{(1)}_{sp^3}$ at site $j$. Where, $<i,j>$ $(<<i,j>>)$ run over all the nearest (next nearest) neighbor hopping sites. The first term represents nearest-neighbor 
hopping in which $t$ is the hopping energy. The second term represents the effective spin-orbit coupling where $\textbf{u}_{ij}=\frac{\textbf{d}_i\times \textbf{d}_j}{\left| \textbf{d}_i\times{\textbf{d}_j}\right| }$, $\textbf{d}_i$ and $\textbf{d}_j$   
denote the nearest bonds that connect the next nearest neighbors. In addition, $u_{ij}=1$ or $u_{ij}=-1$ in case that next-
nearest neighbor hopping is counterclockwise or clockwise respectively. for the 
electrons in the sublattice $A$, $\mu_{ij}=+1$ and for the electrons in the sublattice $B$, $\mu_{ij}=-1$. $t_{SO}$ is the strength of the effective spin-orbit coupling and the Pauli matrix in the spin space is denoted by $\sigma$.  The third term represents the intrinsic Rashba 
spin-orbit coupling related to the next-nearest-neighbor hoping. The forth term 
indicates the external Rashba spin-orbit coupling which is associated with the first-nearest-neighbor hoping. As a consequence of inversion symmetry breaking, this term could be induced by an external electric field or a substrate. The last term is the sublattice potential term which arises from structural buckling. $l$ stands for the buckling height, $\zeta_i=+1(-1)$ for the sublattice $A(B)$ degree of freedom and $E_z$ is the external electric field perpendicular to the 2D  sheet which can control the amount of staggered sublattice potential, $lE_z$.
\begin{table}
	\label{table:structure}
	\caption{Lattice constant and Energy scales for graphene and other buckled honeycomb materials \cite{ezawa1,Min2}.
	}
	\begin{tabular}{clcrcc}
		\hline
		\hline
		\\
		\ \ material \ \   & \ \ $a$(\AA)\ \       & \ \ $t$(eV)\ \     & \ \ $t_{SO}$(meV) \ \   & \ \ $t_{intR}$(meV) & $l$(\AA)\\
		\hline\noalign{\smallskip}
		silicene           &      3.86   & \ \  1.6eV    & \ \   0.75     & \ \ 0.46 & 0.23\\
		\\
		graphene           &    2.46     & \ \   2.8   & \ \ 0.00114    & ---    & 0 \\
		
		\hline
		\hline
	\end{tabular}
\end{table}
\begin{widetext}
The Hamiltonian reads as follow
\begin{eqnarray}
H=\left(
\begin{matrix}
	\eta  & 0 & \gamma _k & i(\lambda_+ + \beta_+) \\
	0 & -\eta  & i(\lambda_-+\beta_-) & \gamma_k \\
	\gamma_k^* & -i(\lambda _-^*+\beta _-^*) & -\eta  & 0 \\
	-i(\lambda _+^*+\beta _+^*) & \gamma_k^* & 0 & \eta  \\
\end{matrix}
\right).
\end{eqnarray}
\end{widetext}
Where we have defined following parameters
\begin{eqnarray}
\eta &=&t_{SO}(2\sin(k_ya)-4\sin(\frac{\sqrt{3}}{2}k_xa)\cos(\frac{k_ya}{2}))\\
\left| \gamma \right| ^2&=&1+4\cos(\frac{\sqrt{3}}{2}k_ya)\cos(\frac{3}{2}k_xa)+4\cos^2(\frac{\sqrt{3}}{2}k_ya)\nonumber
\end{eqnarray}
and also $\lambda _\pm=\lambda _1\pm\lambda _2$, $\beta _\pm=\beta _1\pm\beta _2$ 
\begin{eqnarray}
\lambda _1&=&2it_{intR}(\sin(k_ya)-\sin(\frac{k_ya}{2})\cos(\frac{\sqrt{3}}{2}k_xa))\\
\lambda _2&=&2\sqrt{3}t_{intR}\cos(\frac{k_ya}{2})\cos(\frac{3}{2}k_xa)\\
\beta _1&=&t_{extR}\exp (-i\frac{k_xa}{2\sqrt{3}})\sin(\frac{k_ya}{2})\\
\beta _2&=&\frac{\sqrt{3}}{3}t_{extR}(\exp (i\frac{k_xa}{\sqrt{3}})+\exp (-i\frac{k_xa}{\sqrt{3}})\cos(\frac{k_ya}{2}))\nonumber\\
\end{eqnarray}
To provide a clear understanding of the most important factors in the light induced effects we have ignored the intrinsic Rashba interaction in the numerical computations.

The interaction between the electrons and an electromagnetic 
radiation field in length gauge and long-wavelength   approximation inside the silicene has the form 
\cite{avetissian2013nonlinear}
\begin{eqnarray}
V=e\int d\overset{\rightharpoonup }{r}\hat{\psi}^{\dagger }(\overset{\rightharpoonup }{ {r}}) (\overset{\rightharpoonup }{r}.\overset{\rightharpoonup
}{E}(t) )\hat{\psi} (\overset{\rightharpoonup }{r})
\end{eqnarray}
By expanding the field operators in terms of the silicene wave functions 
\begin{eqnarray}
\hat{\psi} (\overset{\rightharpoonup }{r})&=&\frac{1}{\sqrt{N}}\sum _{k,R_A} e^{-ik.R_A}\phi^{(1)} _{  {sp}^3}\left(r-R_A\right)a_k\nonumber\\&&+\sum _{k,R_B} e^{-ik.R_B}\phi^{(1)} _{  {sp}^3}\left(r-R_B\right)b_k
\end{eqnarray} 
where 
$a^\dagger_k$ ($a_k$) is creation (annihilation) operator which creates (annihilates) an electron with wave vector $k$ in sublattice $A$. Similarly
$b^\dagger_k$ ($b_k$) is creation (annihilation) operator which creates (annihilates) an electron with wave vector $k$ in sublattice $B$. Using above relations one can write
\begin{eqnarray}
V&=&\sum_{k,k'}(\overset{\rightharpoonup }{D}_{  {AA}}.\overset{\rightharpoonup }{E}(t)a^{\dagger }_{k'}a_k+\overset{\rightharpoonup}{D}_{{AB}}.\overset{\rightharpoonup }{E}(t)a^{\dagger }_{k'}b_kf(k)\nonumber\\&&+\overset{\rightharpoonup }{D}_{{BA}}.\overset{\rightharpoonup
}{E}(t)b^{\dagger}_{k'}a_k f^*(k)+\overset{\rightharpoonup}{D}_{{BB}}.\overset{\rightharpoonup}{E}(t)b^{\dagger }_{k'}b_k)\delta _{{k,k}'}\nonumber\\
&=&\left(
\begin{matrix}
\overset{\rightharpoonup }{D}_{\text{AA}}.\overset{\rightharpoonup }{E}(t) & \overset{\rightharpoonup }{D}_{\text{AB}}.\overset{\rightharpoonup
}{E}(t)f(k) \\
\overset{\rightharpoonup }{D}_{\text{BA}}.\overset{\rightharpoonup }{E}(t)f^*(k) & \overset{\rightharpoonup }{D}_{\text{BB}}.\overset{\rightharpoonup
}{E}(t) \\
\end{matrix}
\right)\otimes I_S\nonumber\\
\end{eqnarray}
In which $I_S$ is the identity operator in spin space and electric dipole moment is defined by the following relation
\begin{eqnarray}
\label{(dipole)}
\overset{\rightharpoonup }{D}_{\alpha \beta }&=&\int d^3r\phi^{(1)*} _{\text{sp}^3}(r-R_A)e\overset{\rightharpoonup }{r}\phi^{(1)} _{\text{sp}^3}\left(r-R_B\right),\nonumber\\&&(\alpha ,\beta =A,B)
\end{eqnarray}
and $f(k)=\sum_i e^{i\overset{\rightharpoonup }{k}.\overset{\rightharpoonup }{\delta }_i}$ where $\overset{\rightharpoonup }{\delta }_i$ are the nearest neighbors position vectors. Electric part of the light emission is identified by $\overset{\rightharpoonup}{E}(t)=E(t)\hat{\epsilon}_p$ in which $\hat{\epsilon}_p$ denotes the light polarization and in the present work has been assumed to be  $\hat{\epsilon}_p=(1,i,0)$ for a circularly polarized incident wave.
\\
Eventually, the light-matter interaction takes the form
\begin{eqnarray}
V=\left(
\begin{matrix}
\overset{\rightharpoonup }{D}_{ {AA}} & 0 & f(k)\overset{\rightharpoonup }{D}_{ {AB}} & 0 \\
0 & \overset{\rightharpoonup }{D}_{ {AA}}   & 0 & f(k)\overset{\rightharpoonup }{D}_{ {AB}} \\
f^*(k)\overset{\rightharpoonup }{D}_{ {BA}}   & 0 & \overset{\rightharpoonup }{D}_{ {BB}} & 0 \\
0 &   f^*(k)\overset{\rightharpoonup }{D}_{ {BA}} & 0 & \overset{\rightharpoonup }{D}_{ {BB}} \\
\end{matrix}
\right)\cdot\overset{\rightharpoonup }{E}(t)\nonumber\\ 
\end{eqnarray}
Where according to Eq. (\ref{(dipole)}), it can be shown that the off-diagonal elements  of the dipole moment i.e. $D_{AB}$ and $D_{BA}$ are relatively small
${\overset{\rightharpoonup }{D}_{{AB}}=\overset{\rightharpoonup }{D}_{{BA}}\simeq \left(5.96\times 10^{-5}, -6.13\times 10^{-10}, -5.12\times
10^{-5}\right)}$  (in the unit of electron-Angstrom; $e\AA$) in comparison with diagonal elements, $D_{AA}$ and $D_{BB}$ that have considerable values given by
$\overset{\rightharpoonup }{D}_{{AA}}=\overset{\rightharpoonup }{D}_{{BB}}\simeq\left( 1.915\times 10^{-5}, -4.112\times 10^{-7}, -0.240\right) $ ($e\AA$). 
\\

It should be noted that even small  off-diagonal elements of the dipole moment have to be 
considered. Ignoring these  components leads to a diagonal $V$ matrix which cannot result in inter-band transitions and non-equilibrium polarization injection.

According to Inglot et al \cite{inglot2015m}, the injection rate of any quantity $\hat{O}$ is 
obtained using the well-known Fermi's golden rule,
\begin{eqnarray}
O(\omega )&=&\sum _{n,n'} O^{n\rightarrow n'}(\omega )\\
O^{n\rightarrow n'}  {(\omega )}  &=&  {\frac{2\pi }{\hbar }}  {\int \frac{d^2k}{(2\pi )^2}}\left| \left\langle \psi _{ {nk}}\left|\hat{V}\right|\psi_{n'k}\right\rangle \right| ^2\hat{O}^{n\rightarrow n'}\nonumber\\&&\times\delta (E_{ {nk}}+\hbar \omega -E_{n'k})f(E_{ {nk}})(1-f(E_{n'k}))\nonumber
\end{eqnarray}
Where $n$ and $n'$ are the numbers of sub-bands which optical transitions occur within 
and $f(E_{ {nk}})$ is the Fermi-Dirac distribution function.
Then the non-equilibrium quantities such as spin injection 
\begin{eqnarray}
S_z^{n\rightarrow n'}=\left\langle \psi _{n'k}\left|\hat{s_z}\right|\psi _{n'k}\right\rangle -\left\langle \psi _{ {nk}}\left|\hat{s_z}\right|\psi
	_{ {nk}}\right\rangle, 
\end{eqnarray} 
non-equilibrium normal pseudo-spin 
\begin{eqnarray}
\tau_z^{n\rightarrow n'}=\left\langle \psi _{n'k}\left|\hat{\tau_z}\right|\psi _{n'k}\right\rangle -\left\langle \psi _{ {nk}}\left|\hat{\tau_z}\right|\psi
_{ {nk}}\right\rangle,
\end{eqnarray}
and pseudo-spin current 
\begin{eqnarray}
J_{\tau_z}^{ n\rightarrow n'}=\left\langle \psi _{n'k}\left|\hat{J}^{\tau_z}\right|\psi _{n'k}\right\rangle -\left\langle \psi _{ {nk}}\left|\hat{J}^{\tau_z}\right|\psi _{ {nk}}\right\rangle 
\end{eqnarray}
could be defined within this approach where $\hat{J}^{\tau_z}_{\alpha}= \frac{1}{2}(\hat{v}_\alpha \hat{\tau}_z+\hat{\tau}_z \hat{v}_\alpha)$ and $\hat{v}_\alpha = \frac{1}{\hbar}\frac{\partial \hat{H}}{\partial k_\alpha}$ is the band velocity along the $\alpha$-direction.

$O^{n\rightarrow n'}$ measures the change of expectation value of $O$ via the transition from
initial energy band $n$ to band $n'$ i.e.
$O^{ n\rightarrow n'}=\left\langle \psi _{n'k}\left|\hat{O}\right|\psi _{n'k}\right\rangle -\left\langle \psi _{ {nk}}\left|\hat{O}\right|\psi _{ {nk}}\right\rangle$ . Therefore, induced polarization represents the variation of operator $O$ due to transitions i.e. the non-
equilibrium polarization of system due to optical pumping. 

In the light absorption process, at the range of photon energies that are slightly greater than the band gap, there is another key factor which specifies 
how much a transition could be 
effective in generation of polarization. The electron-hole asymmetry is a consequence the Rashba spin-orbit interaction \cite{shakouri2013effect}. For those transitions that take place between valence and conduction bands particle-hole asymmetry can enhance the non-equilibrium values of a given observable. Asymmetry  between the conduction and valence bands results in more effective transitions in which the change of expectation values for excited electron is high. 
\section{The influence of staggered potential}
In the current investigation, a buckled silicene structure has been considered within 
the semi-classical approach. In the case of graphene, applying a large external 
magnetic field is crucial for observing photogalvanic effect and light induced spin polarization \cite{inglot2015m}. In this case Fermi surface deformation caused by external magnetic field, has been considered to be responsible for non-equilibrium spin-current \cite{inglot2015m}. However, in the case of silicene, significant photon induced polarizations can be injected even at zero magnetic field but in the presence of a normal static electric field. The spin-orbit coupling in silicene is strong enough and can sufficiently lead to band energy splitting, meanwhile, normal electric field results in pseudo-spin polarized bands and spin polarized valleys  close to the Dirac points.

When the perpendicular static electric field is applied to silicene sheet, due to different 
responses of silicon atoms at each of the sublattices, there exists a potential difference between 
two sublattices which is known as \emph{staggered potential}
\cite{lopez2015photoinduced}. Exactly analogous to an external magnetic field 
which can induce a net spin polarization, this static 
perpendicular electric field can  induce pseudo-spin polarization. In this work the 
role of perpendicular static electric field in light induced polarizations has 
been numerically studied. It has been observed that in the absence of this field, there is no non-equilibrium pseudo-spin polarization. By switching on the normal electric field, significant non-equilibrium polarization can be generated. In addition, external electric field remarkably enhances other light induced polarizations such as pseudo-spin current. 

\begin{figure}[h]
	\includegraphics[width=1.0\linewidth]{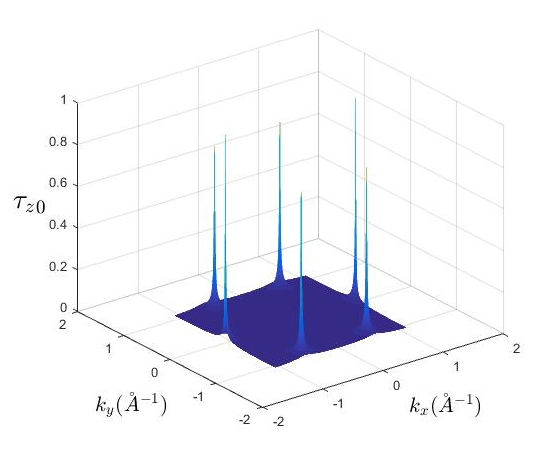}
	\centering
	\caption{(Color online) Pseudo-spin polarization of the first conduction band (n=3). States are fully pseudo-spin polarized very close to the Dirac points and become partially polarized away from these points. \label{PSploarn3}}
\end{figure}

\begin{figure}[h]
	\includegraphics[width=1.0\linewidth]{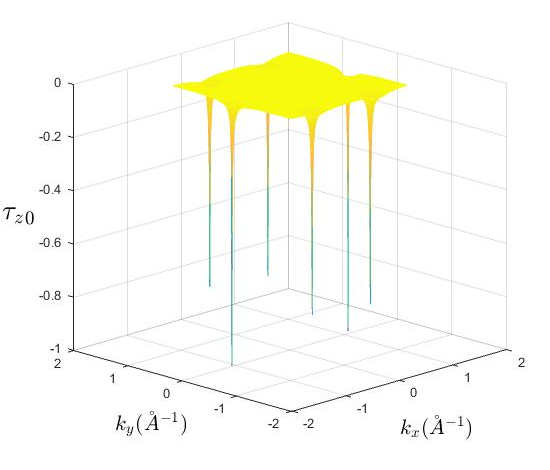}
	\centering
	\caption{(Color online) Pseudo-spin polarization of the second valence  band (n=2).\label{PSploarn2} }
\end{figure}

Applying the perpendicular static electric field which produces 
a staggered potential between two sublattices, breaks the inversion symmetry and subsequently induces a net pseudo-spin polarization at equilibrium (Figs. \ref{PSploarn3} and \ref{PSploarn2}) around the Dirac points. As shown in the Figs. \ref{PSploarn3} and \ref{PSploarn2} each band has a specified normal pseudo-spin polarization at Dirac points where the successive band has opposite pseudo-spin polarization.  However, as it has been shown in this work pseudo-spin polarized bands provide more effective framework for light induced polarizations. Normal electric field is responsible for pseudo-spin polarized bands at equilibrium meanwhile, it should be noted that the non-equilibrium light induced pseudo-spin polarization has also been enhanced remarkably by the staggered filed. 

The main role of the perpendicular electric field can be clearly seen in equilibrium condition i.e. in the absence of radiation field when silicene sheet is exposed to perpendicular Electric field, $E_z$, there will exist a net pseudo-spin equilibrium polarization.  


As it can be seen in the previous figures, in the absence of light emission, applying perpendicular electric field solely can induce a net pseudo-spin polarization in both valence and conduction bands. Normal component of pseudo-spin, $\tau_z$, is a good quantum number close to the Dirac points. Therefore, it is noticeable that any factor which causes an inter-band transition, this band change is also accompanied with a pseudo-spin flip of electron. 

\begin{figure}[!]
	\includegraphics[width=1.0\linewidth]{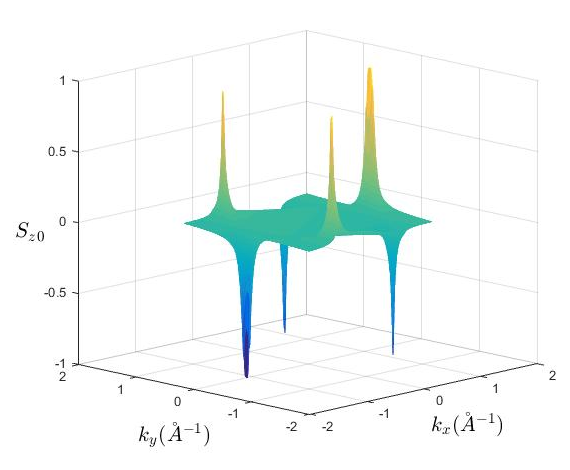}
	\centering
	\caption{(Color online) Spin polarization of the second valance band (n=2).\label{Sploarn2}}
\end{figure}
\begin{figure}[!]
	\includegraphics[width=1.0\linewidth]{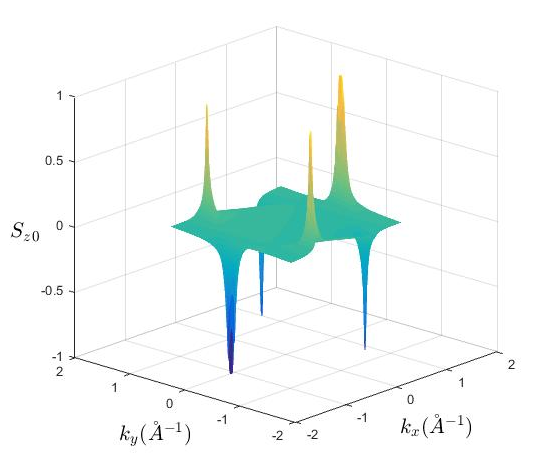}
	\centering
	\caption{(Color online) Spin polarization of the first conduction band (n=3).\label{Sploarn3}}
\end{figure}

On the other hand, as mentioned earlier normal electric field results in spin-valley locking. This can be inferred from the Figs. \ref{Sploarn2} and \ref{Sploarn3} the this effect is almost band-independent. This normal spin preserving property of electrons at each of the valleys has been removed by lifting the normal electric field. 
\\

The pseudo-spin polarized bands and spin-valley locking which has been observed at vicinity states of the Dirac points can be explained simply within the Dirac point approximation.
given that extrinsic Rashba coupling is small (that is about $0.001$eV), in the presence of perpendicular electric field and very close to the Dirac points ($k\rightarrow 0$), low energy Dirac Hamiltonian is reduced as 
$H = diag\{\eta \lambda _{  {SO}}+  {lE}_z,-\eta \lambda _{  {SO}}+  {lE}_z,  -\eta \lambda _{  {SO}}-  {lE}_z, \eta \lambda _{  {SO}}-  {lE}_z  \}$. Accordingly, in this case the Hamiltonian is diagonal in the bases of $|\tau_z>\otimes|\sigma_z>$ which means that the eigenstates have definite pseudo-spin and spin quantum numbers. Eigenvalue of this eigenstate is given by $\varepsilon^\eta_{\tau\sigma}= \eta\sigma_z \lambda _{  {SO}}+  \tau_z{lE}_z$ where $\varepsilon^\eta_{\tau\sigma}=\varepsilon^{-\eta}_{\tau-\sigma}$ clearly exhibits spin-valley locking that electrons of the same band and different valleys are isoenergetic when their spins directed oppositely. Meanwhile, there is a gap between the pseudo-spin up and pseudo-spin down states given by $\Delta=2|lE_z-\lambda_{SO}|$ which indicates that at zero electric field ($E_z=0$) the energy difference between up and down spins is very small of order of spin-orbit coupling and therefore population difference between the opposite spins is negligible. These results directly reflect the time-reversal symmetry of the system when the inversion symmetry is broken.
In addition this electrically tunable band gap provides a topological phase transition by manipulation of the normal electric field \cite{ezawa2013m,ezawa2015monolayer}.

As it is obviously shown up to here, perpendicular electric in silicene represents rich physics from topological phase transition to pseudo-spin band polarization and spin-valley locking \cite{ezawa2013m,ezawa2015monolayer,tao2019two}. 
\section{Valley polarization: generation and detection }
Valley polarization can be generated in different ways in graphene-like materials. Magnetic barrier could be inserted in a strained graphene sheet as a valley filter. By adjusting the strength of strain and including a scalar electric potential, amplitude of valley polarization can be tuned meanwhile, its polarity is controllable by switching the direction of the local magnetic field \cite{zhai2010magnetic}.

Since one can associate an intrinsic magnetic moment to valley index, it is possible to measure this degree of freedom directly. This valley-contrasting magnetic moment is called orbital magnetic moment and originates from self-rotation of Bloch wave packets. Thus coupling between orbital magnetic moment and an external magnetic field can produce valley polarization \cite{xiao2007valley}.
\begin{figure*}[t]
	\includegraphics[width=0.8\linewidth]{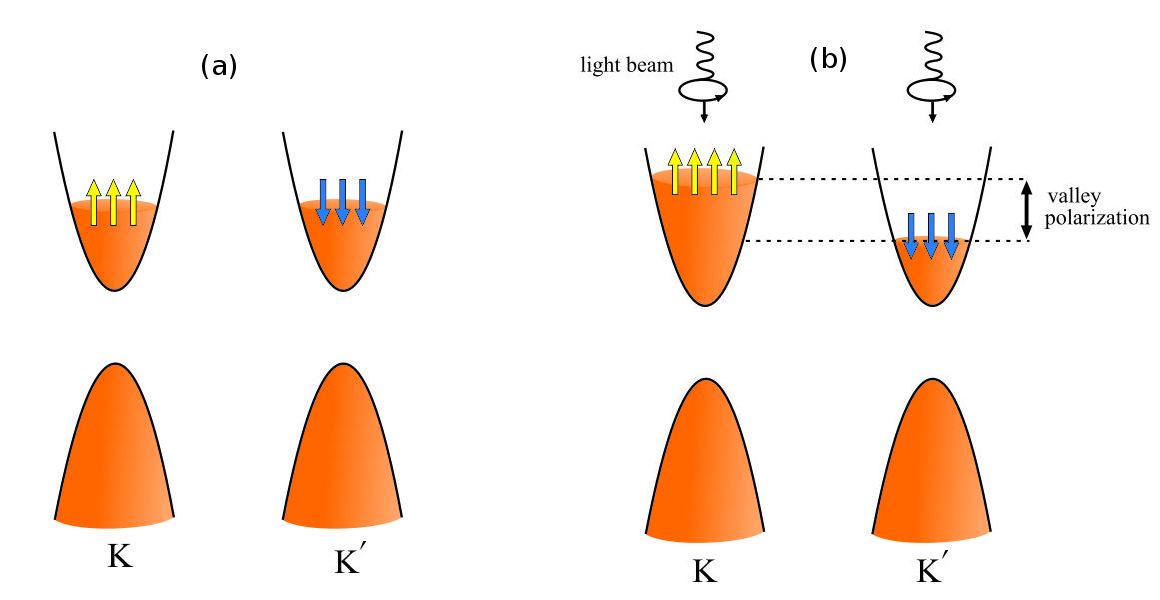}
	\centering
	\caption{(Color online) (a) Normal field induced spin-valley locking. (b) Non-symmetric absorption of the circularly polarized light \cite{ezawa2012spin} that generates spin and valley polarization.\label{Vpolar}}
\end{figure*}
There have been introduced several detection methods of valley polarization. Di Xiao et al have  measured this quantity as a signal of orbital magnetization. Due to a quantized valley-dependent Hall conductance, applying an external in-plane electric field induces a net Hall current which in turn results in a transverse measurable voltage. Therefore the measurement of this voltage can be interpreted as detection of valley polarization \cite{xiao2007valley}. Moreover, it has been shown that by utilizing a universal connection among optical oscillator strength of inter-band transitions, the orbital magnetic moment and the Berry curvature, one can optically measure the orbital magnetization and intrinsic anomalous Hall conductivity in ferromagnetic systems \cite{yao2008valley}. The breaking of inversion symmetry results in opposite Berry curvatures in two valleys. In the absence of external magnetic field, the net Hall current does not exist but applying a circularly polarized light induces additional populations of electron and holes which their travel to opposite edges of sample generates a transverse voltage. The sign of this voltage indicates the polarization of incident light \cite{yao2008valley}.

There is another method that has been suggested for measurement of valley-polarized current based on transport features in a T-junction of two ribbons which was reported in graphene systems \cite{chan2018detection}. The measured currents in the armchair and zigzag leads are used to deduce the valley polarization of incoming current \cite{chan2018detection}. 

In the present study it has been shown that how spin-valley locking provides a simple magnetic method for detection of the valley polarization. This is originally based on the fact that non-equilibrium spin population and valley population imbalance are proportional to each other as a result of the spin-valley locking. 
\\
\section{ discussion and results}
\subsection{Light-induced spin and valley polarization}
In the presence of staggered potential and due to spin-valley locking, spin population imbalance is almost identical with the valley polarization at low Fermi energies. In other words, since non-equilibrium spin polarization coincides with the population imbalance between the two valleys, therefore spin polarization measures indirectly the valley polarization. 

\begin{figure}[h]
	\includegraphics[width=1.0\linewidth]{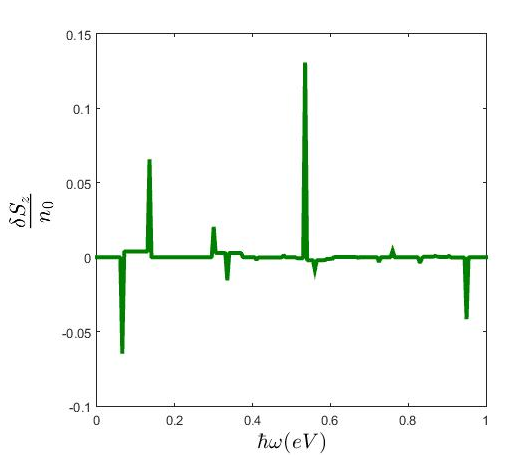}
	\centering
	\caption{(Color online) Spin polarization as function of the photon frequency when $lE_z=0.03eV$.\label{Sz_hw}}
\end{figure}
At first look it seems that the spin-valley locking results in exactly identical population of $s_z$-spins with its corresponding valley population. However, it is should be noted that all of the occupied states in a single valley are not completely $s_z$ polarized. Equilibrium spin polarization of a given state depends on the position of this state in $k$-space. Very close to the Dirac points states are fully spin polarized and as the distance of the state from the Dirac point increases, spin polarization of the state decreases (Figs \ref{PSploarn2} and \ref{PSploarn3}). At low Fermi energies the spin and valley populations are very close while away from the Dirac points states are partially polarized.        
Assuming that the up (down) spins are located around the $K$ ($K'$) valley (as a result of the spin-valley locking) the relation between the spin and valley population at each of the valleys can be given by the following expressions,
\begin{eqnarray}
n_\uparrow &=& \alpha_K n_K\nonumber\\
n_\downarrow &=& \alpha_{K'} n_{K'}, 
\end{eqnarray}
where, $n_K$ ($n_{K'}$) stands for the valley population of $K$ ($K'$) valley and $n_\uparrow$ ($n_\downarrow$) is the spin population of the same valley. $\alpha_\eta$ is a coefficient that measures the ratio of the polarized state in the $\eta$-valley which depends on the Fermi wave number ($k_F$). Due to the symmetry of the valleys one can expect that $\alpha_{K}=\alpha_{K'} =\alpha$ and accordingly one can easily obtain $Ps = \alpha~ Pv $ where $Ps=(n_\uparrow-n_\downarrow)/n$ and  $Ps=(n_K-n_{K'})/n$ are the spin polarization and valley polarization of the system respectively. Meanwhile, the $n =n_K+n_{K'}$ is the electronic population of the sample.  
$\alpha$ can simply be estimated by the following relation,  
\begin{eqnarray}
 \alpha_K=\frac{1}{(\hbar/2) n}<Sz>_K, 
\end{eqnarray}   
in which $<Sz>_K$ is sum of the expectation value of $\hat{s}_z$ over all of the occupied states in a single valley. Therefore, spin and valley polarization can be related to each other via $Ps = \alpha~ Pv $.    
\\
As it is shown in the Fig. \ref{Vpolar} in the absence of irradiation there is no net spin polarization. After applying the circularly polarized electromagnetic field, as the charge carriers in two different valleys, $K$ and $K'$, response differently to this specific chirality of light, there will emerge different population of electrons in each of valleys which is known as valley polarization (Fig. \ref{Vpolar}).
\\

Behavior of the system depends on the  chirality of circularly polarized incident light. When the silicene sheet is exposed to  circular right-handed polarized light, the behaviour of spin in $K$ valley is exactly the same as $K'$ valley in the presence of left-handed light and vice-versa. Accordingly, it is expected that each of these circularly polarized lights could provide one of the valley polarizations in the silicene \cite{ezawa2012spin}.
\\

Numerical results which have been represented  here are normalized to electrons density of the system, $n_0\simeq 1.5\times 10^{15}cm^{-2}$ and temperature given in the distribution functions has been chosen to be $T=1$K.
\\

\begin{figure}[h]
	\includegraphics[width=1.0\linewidth]{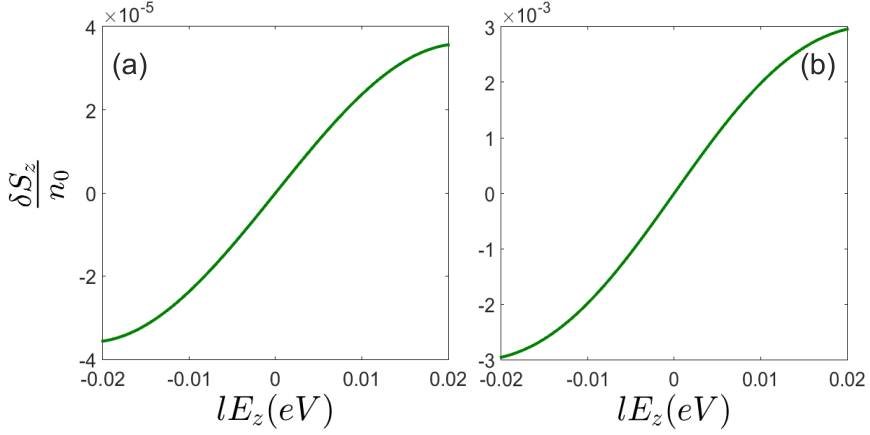}
	\centering
	\caption{(Color online) Spin polarization of (a); $K$ and (b); $K'$ valleys in terms of the normal electric field at $\hbar\omega =0.3eV$. \label{Sz_Ez}}
\end{figure}

As shown in Fig. \ref{Sz_hw} circularly polarized light can generate spin polarization. At several frequencies in which the band energy difference coincides with the energy of absorbed photon spin-polarization shows remarkable jumps. Meanwhile, as indicated in Fig. \ref{Sz_Ez} spin polarization is suppressed at $E_z = 0$. This can be explained by the concept of the spin-valley locking and non-symmetric light absorption of the left and right-handed photons. As discussed earlier, light induced spin polarization in silicene can be considered as a direct consequence of the spin-valley locking that has been induced by the normal external electric field.      
\\

Since momentum transfer of photon induced transitions is small, this type of state-change cannot be responsible for inter-valley transitions.   
The momentum difference between two valleys is of the order of $\Delta k\sim \left| K-K'\right|$ which is quite large amount in the momentum space for photon induced transitions at energy range where the light absorption rate is not zero \cite{xiao2007valley}. Meanwhile some short-range impurities can provide such a large momentum transfer. It should be reminded that the calculations have been performed beyond the Dirac point approximation where any possible inter-valley transition can be captured within this method. 

\begin{figure}[h]
	\includegraphics[width=0.70\linewidth]{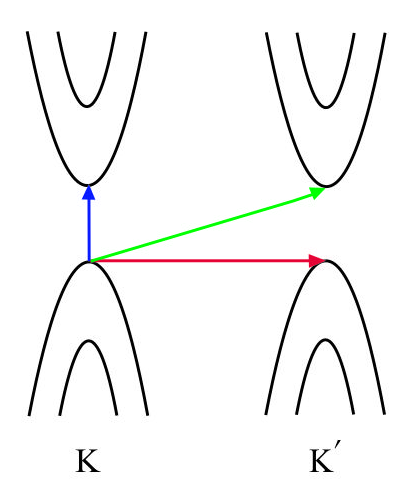}
	\centering
	\caption{(Color online) Blue arrow indicates intera-valley inter-band transition. Red arrow represents intra-band inter-valley elastic transition and green arrow is a schematic example of inter-valley inter-band transition. \label{transition}}
\end{figure}

In the regime of elastic scatterings of impurities, for insulating silicene there is no inter-band induced relaxation thus electron population at each of the valleys remains unchanged. Different type of transitions has been illustrated schematically in Fig. \ref{transition}.
\begin{figure*}[t]
	\includegraphics[width=0.70\linewidth]{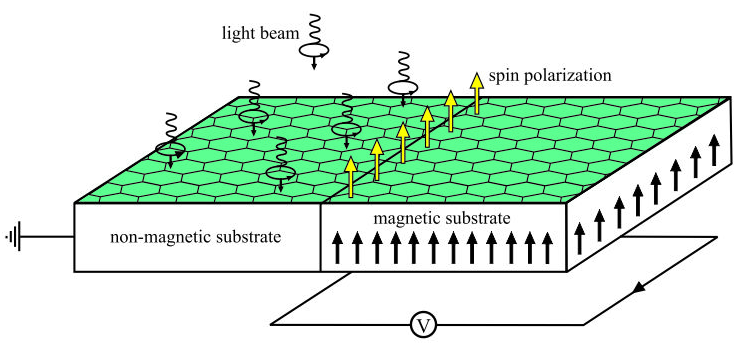}
	\centering
	\caption{(Color online) Spin based detection of the valley polarization  in which the light induced valley polarization could be measured by its corresponding spin polarized population.\label{setup} }
\end{figure*}
Inter-valley transitions which could be originated from the short range impurities can provide the inter-valley momentum transfers. In addition this type of transitions take place when the impurity potential provides a spin flip of electrons since the different valleys are oppositely spin polarized as a result of spin-valley locking. Therefore an inter-valley transition should be followed by a flip of normal spin component. Therefore, it can be inferred that inter-valley transitions in staggered potential regime just could be provided by sharp (short range) and magnetic impurities. 
\\

Spin and valley relaxation times ($\tau_s$ and $\tau_v$) are given by
$\tau _{  {s}} \simeq \tau _{ {v}}\sim \frac{1}{\sum _{K'\lambda '} W_{\lambda \lambda '}^{\xi \xi '}\left(k,k'\right)}$. Where $W_{\lambda \lambda '}^{\xi \xi '}\left(k,k'\right)$  is the transition rate from  $|k\lambda\rangle$ state of $\zeta$ valley to $|k'\lambda'\rangle$ of $\zeta'$.
At the elastic regime with nonmagnetic impurities  one can realize that, $\frac{1}{\tau_s}\simeq\frac{1}{\tau_v}\sim \frac{2\pi v_0^2}{\hbar E_f}n|\overline{\langle u^{\dagger}_{K\lambda}|u_{K'\lambda }\rangle}|^2$. Regardless of the chosen form of electron interaction with light and impurity, in the elastic regime of relaxations, if the short-range 
impurities are nonmagnetic, the probability of inter-valley transitions for the states in 
vicinity of Dirac points is practically negligible $<u^{\dagger }_{K\lambda }u_{K'\lambda }>\rightarrow 0$ where $K$ and $K'$ are valley indices and $\lambda$ stands for energy band (band index should not change in the elastic scatterings therefore both states $u_{K\lambda }$ and $u_{K'\lambda }$ have the same band number). Thus the electron states could be considered as \emph{long-
live spin-valley} states i.e. spin and valley lifetimes are of the same order and large enough to preserve against the relaxations. It should be noted that long-wavelength electromagnetic waves, as considered in this work, cannot provide momentum transfer of inter-valley transitions.
\\


\subsection{Spin-valley locking based Detection of valley-polarization }
Since, the valley population imbalance is proportional to spin polarization it seems that the spin polarization detection devices could be proposed for valley polarization measurements as well. As shown in the Fig. \ref{setup} detection setup includes silicene sheet deposited on a bipartite substrate. One half is non-magnetic and the other half is magnetic. The circuit is initially biased to sweep the carriers from non-magnetic part to the magnetic part which acts as a detection region. 
\\

By applying the electromagnetic radiation on non-magnetic part, if the light-induced magnetic polarization is aligned with magnetization of ferromagnetic substrate, the voltage in circuit will be increased as well as an opposite $\delta M$ leads to voltage decrease.
The net magnetic polarization induced by photoelectric effect reads as follow ${  {\delta M}=\sum \frac{  {\delta S}}{L^3}}$

Where the summation is over the whole volume of the detection region and $L^3$
is the volume of bulk. Then
${V\sim \frac{  {\delta \Phi}}{  {\delta t}}}$, $V$ is the induced voltage and $\Phi$ is the magnetic flow passing through the circuit cross section area, $A$, due to induced current which is related to the normal component of magnetization, $M_n$ by $\Phi =4 \pi A M_n$ \cite{bloch1946nuclear}.  

As discussed before and also has been reported in previous works \cite{xu2014spin,tao2019two}) spin-valley locking enhances spin life-time. Therefore, in the spin-valley locking regime spin relaxation time could be high enough to be effectively detected in the magnetic region. 

\subsection{Non-equilibrium pseudo-spin polarization and pseudo-spin current}
\begin{figure}[h]
	\includegraphics[width=1.0\linewidth]{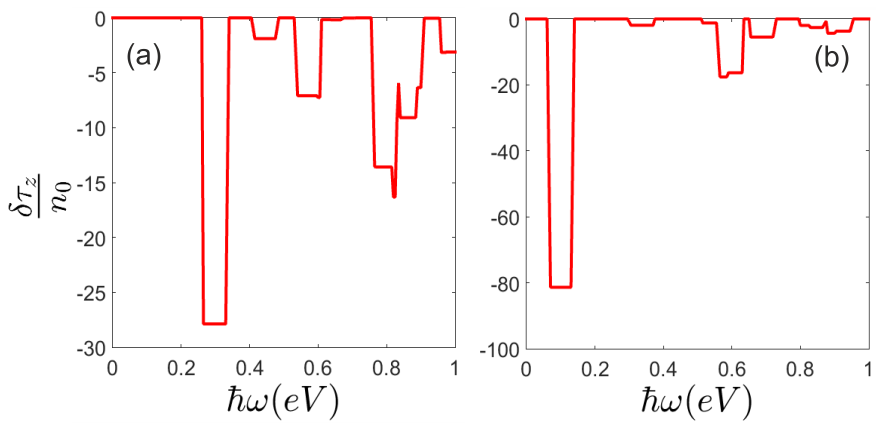}
	\centering
	\caption{(Color online) Pseudo-spin polarization of (a); $K$ and (b); $K'$ valleys at $lE_z=0.03eV$.\label{tau_hw}}
\end{figure}
\begin{figure}[h]
	\includegraphics[width=1.0\linewidth]{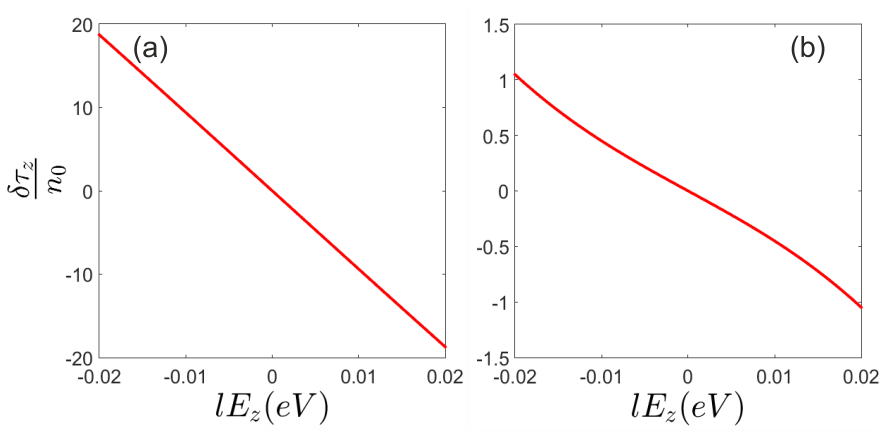}
	\centering
	\caption{(Color online) Pseudo-spin polarization of (a); $K$ and (b); $K'$ valleys in terms of the normal electric field at 
		$\hbar\omega =0.3eV$.\label{tau_Ez}}
\end{figure}
As it can be inferred from Figs. \ref{tau_hw} and \ref{tau_Ez} non-equilibrium pseudo-spin polarization can be induced by circularly polarized photons. The light-matter interaction leads to a considerable non-equilibrium pseudo-spin polarization when the normal static filed is applied. It should be noted that in the absence of perpendicular electric field, non-equilibrium pseudo-spin polarization cannot be injected by the radiation field (Fig. \ref{tau_Ez}). At zero electric field and very close to the Dirac points band energies are given by  $\varepsilon^\eta_{\tau\sigma}= \eta\sigma_z \lambda _{  {SO}}$ where the eigenstates are $|\tau_z>\otimes|\sigma_z>$ and therefore pseudo-spin up and down states located near to these points become degenerate at $E_z =0$. This gives rise to zero pseudo-spin polarization. On the other hand states that have been placed far from the Dirac points are not pseudo-spin polarized and light induced transitions between these states could not result in effective non-equilibrium pseudo-spin polarization.
\\

Meanwhile, since successive pseudo-spin dependent bands in the presence of normal field have opposite sign of pseudo-spin polarization, light induced transitions between these bands result in negative non-equilibrium pseudo-spin polarization (Fig. \ref{tau_Ez}). 

\begin{figure}[h]
	\includegraphics[width=1.0\linewidth]{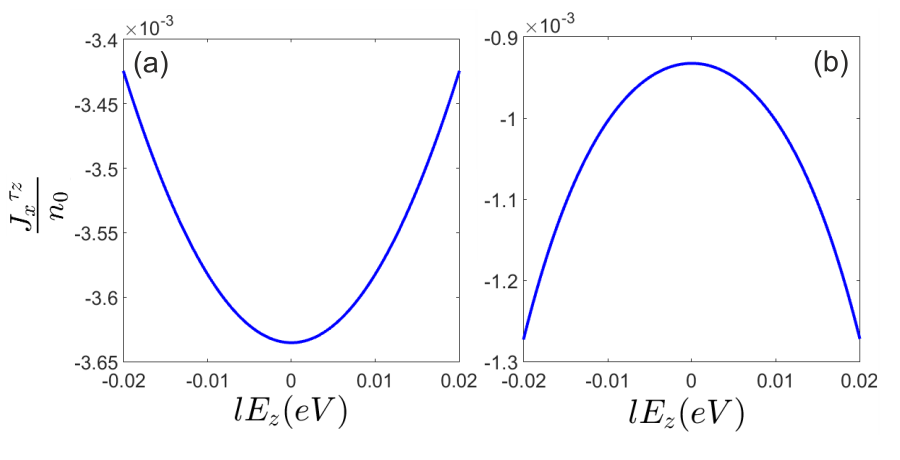}
	\centering
	\caption{(Color online) Light induced pseudo-spin current of (a); $K$ and (b); $K'$ valleys along the  $x$-axis in terms of the normal electric field at $\hbar\omega =0.3eV$. .\label{Jxz}}
\end{figure}

\begin{figure}[h]
	\includegraphics[width=1.0\linewidth]{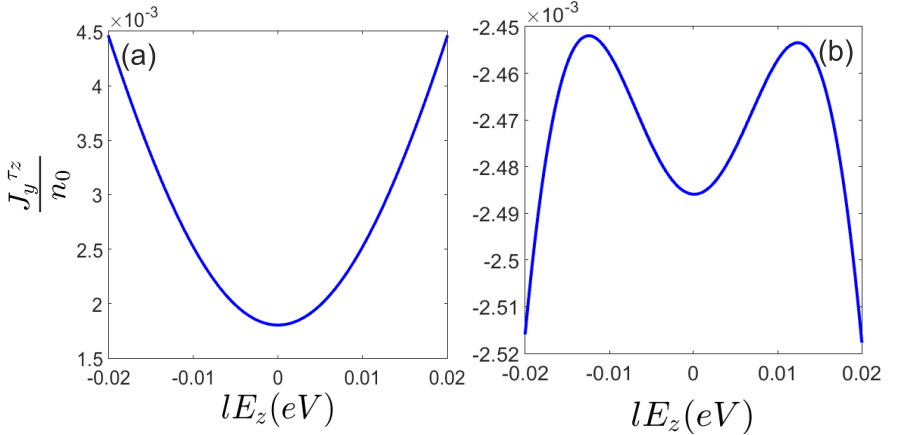}
	\centering
	\caption{(Color online) Light induced pseudo-spin current of (a); $K$ and (b); $K'$ valleys along the $y$-axis in terms of the normal electric field at $\hbar\omega =0.3eV$.\label{Jyz}}
\end{figure}

As depicted in Figs. \ref{Jxz} and \ref{Jyz} light induced transitions can result in non-equilibrium pseudo-spin current in silicene.
Figures show that pseudo-spin-polarized currents are different in two perpendicular directions $x$ and $y$. This means that optical radiation not only injects valley polarization in silicene sheet but also results in anisotropic response of the system. This anisotropy would rely on the silicene band anisotropy which can be captured beyond the Dirac point approximation as performed in the present study. Within this approach band anisotropy manifests itself in the numerical results. 
\\

Since the circularly polarized photon absorption is not symmetric at different valleys, the  population imbalance between the $K$ and $K'=-K$ results in electric current. Therefore
pseudo-spin polarized current increases by normal electric field based on the background pseudo-spin polarization increment. 
\\

However, it is really remarkable that unlike the non-equilibrium spin and pseudo-spin polarizations, pseudo-spin current could be generated even at zero normal electric field (Figs. \ref{Jxz} and \ref{Jyz}). This reflects the fact that not only the light absorption of different valleys are not symmetric but also this absorption is not identical for each of the states around a single Dirac point. On the other words, in the absence of the staggered  potential each of the band states has different pseudo-spin polarization, $<\tau_z>$, depending on the location of the state in $k$-space and band number, however, the total pseudo-spin polarization of the band is zero. When the circularly polarized light excites non-symmetrical states around the Dirac point overall induced pseudo-spin current becomes nonzero.
\\
       
Meanwhile, this can be considered as a specific consequence of the present approach which goes beyond the Dirac point approximation. Within the Dirac point approximation pseudo-spin current operator is given by $J^{\tau_z}_{x(y)}=v_F\frac{1}{2}(\tau_z\tau_{x(y)}+\tau_{x(y)}\tau_z)$ (when the internal Rashba interaction has been ignored) that identically vanishes which shows that non-equilibrium pseudo-spin current cannot be captured within the Dirac point approximation. 
\section{Concluding Remarks}
Within the semi-classical approach light-induced polarization in silicene has been investigated beyond the Dirac point approximation.
Calculations are performed at the long-wavelength limit wherein the amplitude of external field remains practically constant over the atomic scale, where light-matter interaction at this limit cannot lead to inter-valley transition.

It has been shown that the normal electric field results in light induced spin and pseudo-spin polarization. Meanwhile, this field enhances the light induced pseudo-spin current.

Non-symmetric absorption of circularly polarized light are known to be responsible for spin polarization and pseudo-spin current. 

Meanwhile, it is noteworthy, since inside each valley initial and final states of transition possess different pseudo-spins. Therefore, light induced transitions from one energy band to another leads to pseudo-spin flip i.e. negative non-equilibrium pseudo-spin polarization.

Spin-valley locking provides a magnetic framework for detection of the valley polarization by non-equilibrium spin measurements.

\begin{acknowledgements}
	We are very much grateful to Negin Shahabi for embodying our scientific imaginations in the form of well-favoured pictures.
\end{acknowledgements}
\appendix*
\section{}
\label{app}
Spherical harmonics read as
\begin{eqnarray}
\label{Ylm}
Y_{00}(\theta ,\varphi )&=&\frac{1}{\sqrt{4\pi}}\nonumber\\
Y_{10}(\theta ,\varphi )&=&\sqrt{\frac{3}{4  \pi}}  \cos\theta \\
Y_{1,\pm 1}(\theta ,\varphi )&=&\sqrt{\frac{3}{8  \pi}}{\sin}\theta e^{\pm \text{i$\varphi $}}.\nonumber
\end{eqnarray}
Where $s$ and $p$ orbitals can be written as
\begin{eqnarray}
s(\theta ,\varphi )&=&Y_{00}(\theta ,\varphi )\\
p_x(\theta ,\varphi )&=&\frac{1}{\sqrt{2}}\left\{Y_{1,1}(\theta ,\varphi )+Y_{1,-1}(\theta ,\varphi )\right\}\nonumber\\
p_y(\theta ,\varphi )&=&\frac{1}{\sqrt{2}}\left\{Y_{1,1}(\theta ,\varphi )-Y_{1,-1}(\theta ,\varphi )\right\}\nonumber\\
p_z(\theta ,\varphi )&=&Y_{10}(\theta ,\varphi )\nonumber
\end{eqnarray}
Meanwhile, following radial wave functions have also been employed for defining the hybrid orbitals:
\begin{eqnarray}
\label{Rnl}
R_{30}(r)&=&\frac{2}{\sqrt{27}}\left(\frac{a}{Z}\right)^{\frac{-3}{2}}\left(1-\frac{2}{3}\left(\frac{ {Zr}}{a}\right)+\frac{2}{27}\left(\frac{ {Zr}}{a}\right)^2\right)e^{-\left(\frac{ {Zr}}{3a}\right)}\nonumber\\
R_{31}(r)&=&\frac{8}{27\sqrt{6}}\left(\frac{a}{Z}\right)^{\frac{-3}{2}}\left(1-\frac{1}{6}\left(\frac{{Zr}}{a}\right)\right)\left(\frac{{Zr}}{a}\right)e^{-\left(\frac{{Zr}}{3a}\right)}\nonumber\\
\end{eqnarray} 

\bibliographystyle{apsrev4-1}
\bibliography{refrences}
\end{document}